\documentclass[journal=jacsat,10pt,manuscript=article,layout=twocolumn]{achemso}

\usepackage{mathtools, cuted}
\usepackage{caption}

\usepackage{lineno,hyperref}
\modulolinenumbers[5]
\usepackage{amssymb}
\usepackage{times}
\usepackage{graphicx}
\usepackage[usenames,dvipsnames,svgnames,table]{xcolor}
\usepackage{natbib}
\usepackage{array}
\usepackage{tabularx}
\onecolumn

\author{B.~G.~M\'{a}rkus}
\affiliation{Department of Physics, Budapest University of Technology and Economics and MTA-BME Lend\"{u}let Spintronics Research Group (PROSPIN), PO Box 91, H-1521 Budapest, Hungary}
\alsoaffiliation{Laboratory of Physics of Complex Matter, \'{E}cole Polytechnique F\'{e}d\'{e}rale de Lausanne, Lausanne CH-1015, Switzerland}

\author{P.~Szirmai}
\affiliation{Laboratory of Physics of Complex Matter, \'{E}cole Polytechnique F\'{e}d\'{e}rale de Lausanne, Lausanne CH-1015, Switzerland}

\author{K.~F.~Edelthalhammer}

\author{P.~Eckerlein}

\author{A.~Hirsch}

\author{F.~Hauke}
\affiliation{Department of Chemistry and Pharmacy and Institute of Advanced Materials and Processes
(ZMP), University of Erlangen-N\"{u}rnberg, Nikolaus-Fiebiger-Strasse 10, 91058 Erlangen, Germany}

\author{N.~M.~Nemes}
\affiliation{GFMC, Unidad Asociada ICMM-CSIC "Laboratorio de Heteroestructuras con Aplicacion en Espintronica", Departamento de Fisica de Materiales Universidad Complutense de Madrid, 28040 Madrid, Spain}

\author{J.~C.~Chac\'{o}n-Torres}
\affiliation{Universidad UTE, Facultad de Ciencias, Ingenier\'{i}a y Construcci\'{o}n, 170147 Quito, Ecuador and\\ Yachay Tech~University, School of Physical Sciences and Nanotechnology, 100119 Urcuqu\'{i}, Ecuador}

\author{B.~N\'{a}fr\'{a}di}

\author{L.~Forr\'{o}}
\affiliation{Laboratory of Physics of Complex Matter, \'{E}cole Polytechnique F\'{e}d\'{e}rale de Lausanne, Lausanne CH-1015, Switzerland}

\author{T.~Pichler}
\affiliation{Faculty of Physics, University of Vienna, Strudlhofgasse 4., Vienna, A-1090, Austria}

\author{F.~Simon}
\email{f.simon@eik.bme.hu}
\affiliation{Department of Physics, Budapest University of Technology and Economics and MTA-BME Lend\"{u}let Spintronics Research Group (PROSPIN), PO Box 91, H-1521 Budapest, Hungary}
\alsoaffiliation{Laboratory of Physics of Complex Matter, \'{E}cole Polytechnique F\'{e}d\'{e}rale de Lausanne, Lausanne CH-1015, Switzerland}

\title{Ultralong Spin Lifetime in Light Alkali Atom Doped Graphene}

\begin{document}

\begin{abstract}
Today’s great challenges of energy and informational technologies are addressed with a singular compound, the Li and Na doped few layer graphene. All what is impossible for graphite (homogeneous and high level Na doping), and unstable for single layer graphene, works very well for this structure. The transformation of the Raman G line to a Fano lineshape and the emergence of strong, metallic-like electron spin resonance (ESR) modes, attest the high level of graphene doping in liquid ammonia for both kinds of alkali atoms. The spin-relaxation time in our materials, deduced from the ESR line-width, is 6-8 ns, which is comparable to the longest values found in spin-transport experiments on ultrahigh mobility graphene flakes. This could qualify our material as promising candidate in spintronics devices. On the other hand, the successful sodium doping, this being a highly abundant metal, could be an encouraging alternative to lithium batteries.
\end{abstract}

\section*{KEYWORDS:}{few-layer graphene, alkali atom doping, Raman spectroscopy, electron spin lifetime, electron spin resonance}

\maketitle
\clearpage
\newpage
\twocolumn


Charge state in various forms of carbon can be conveniently controlled using alkali atom doping methods. It led to applications in \textit{e.g.} energy storage \cite{KorthauerLi2017} and to the discovery of compelling correlated phases such as \textit{e.g.} superconductivity (with $T_{\mathrm{c}}=11.5$ K in CaC$_6$ Ref. \cite{EmeryPRL2005,EmerySTAM2008} and $T_{\mathrm{c}}=28$ K in Rb$_3$C$_{60}$ Ref. \cite{RosseinskyPRL1991}), spin density waves in fullerides\cite{JanossyPRL1997}, and the Tomonaga--Luttinger to Fermi liquid crossover in single-wall carbon nanotubes \cite{IshiiNat2003,RaufPRL2004}.

Control over the number of free electrons in carbon could be also exploited in spintronics \cite{ZuticRMP2004,WolfSCI,WuReview}, \textit{i.e.} when using the electron spin as an information carrier and storage unit. Doping-induced metallic carbon has the smallest spin-orbit coupling besides metallic Li and Be, which is expected to result in ultralong spin lifetime, $\tau_{\mathrm{s}}$; a prerequisite for spintronics. Graphene arose as a promising candidate for spintronics purposes  due to the predicted long $\tau_{\mathrm{s}}$, its actual value is controversial according to spin transport studies \cite{TombrosNAT2007,KawakamiPRL2011,OzyilmazPRL2011,OzyilmazNL2011,HanNL2012,DrogelerNL2016} and it ranges from $100$ ps \cite{TombrosNAT2007} up to $12$ ns \cite{KawakamiPRL2011,HanNL2012,DrogelerNL2016} with theoretical hints that the short lifetime originates from extrinsic effects \cite{FabianPRL2015}.
Doping graphene with light alkali atoms would {\color{black}enable the accurate determination} $\tau_{\mathrm{s}}$ for the itinerant electrons by spin spectroscopy, \textit{i.e.} electron spin resonance (ESR) \cite{FeherPR1955,DysonPR1955}. Light alkali atoms have a small spin-orbit coupling \cite{ElliottPR1954,YafetSSP1963} thus the intrinsic spin lifetime in graphene is expected to be observed with this approach.

Conventional alkali doping of carbon proceeds in the so-called vapor phase which works well for heavier alkali atoms with a lower melting point (K, Rb, and Cs) \cite{DresselhausAP1981}: the alkali atoms are heated together with the desired form of carbon (graphite, graphene, nanotubes, or fullerene). For Li and Ca doping, graphite doping was achieved by immersing the sample into molten Li or Li/Ca mixtures \cite{EmeryPRL2005} with temperatures up to $350~^{\circ}$C. This relatively high temperature is dictated by the melting point and is due to the slow kinetics of the diffusion process but the reaction has a small temperature window due to formation of alkali carbides around $450~^{\circ}$C. This method can only be formed for bulky sample (\textit{e.g.} for a piece of HOPG or graphite single crystal) which can be inserted and removed from the molten metal.

Alkali atom doping of graphene is also intensively studied \cite{JungACSNano2011,HowardPRB2011,KumarACSNano2011,XueJACS2012,ParretACSNano2013,ZhouNanoscale2017,PervanPRM2017}. Achieving controllable and high level doping would be particularly important for chemically exfoliated graphene, which is the bulk form of graphene, \textit{e.g.} for energy storage purposes \cite{WangCarbon2013,YangSci2013,MahmoodJMCA2014}. Especially Na doped graphene is an important candidate to replace conventional Li-ion batteries due to the low cost and high abundance of sodium\cite{MedardePRL2013,BarpandaNatComm2014,AresAdvMat2017,TianACSNano2018}.

 However, Na is somewhat of an outlier among the alkali metals as it does not intercalate graphite under ambient pressure except for a very low stoichiometry of NaC$_{64}$\cite{AsherNat1958,AsherJINC1959,MetrotSM1980,AkuzavaMCLC2002,Adelhelm_why_NaGIC_unstable} as compared to \textit{e.g.} LiC$_6$ and KC$_8$. 
 Successful Na doping was reported for graphene, prepared by a solvothermal synthesis by the Choucair method \cite{ChoucairNatMat}, either using electrochemical doping \cite{NaGraphene_ACS_Applied} or when the material was synthesized from Na containing precursors \cite{NaGraphene_AdvScie}. However, solvothermal-derived graphene is known to have a relatively large defect concentration as well as a limited flake size, it is thus intriguing whether Na doping could be achieved for a high quality and high area mono or few-layer graphene.
 

As mentioned, doping with light alkali atoms cannot be performed with the vapor phase method. An alternative doping route uses solvents, such as liquid ammonia and organic solvents (\textit{e.g.} tetrahydrofuran (THF) \cite{BeguinMSE1979} or $2$-methyltetrahydrofuran (2-MeTHF) and $2,5$-dimethyltetrahydrofuran (diMeTHF) \cite{MizutaniJPCS1996,MizutaniSM2001}). These are known to dissolve well the alkali and some alkaline-earth elements. Then, the reaction between the alkali atoms and the carbon material proceeds in the solution at moderate temperatures and with a high efficiency due to the large reaction surface. This procedure was used to obtain highly doped fullerides \cite{MurphyJPCS1992,BuffingerJACS1993,DahlkeJACS2000,AjayanACSNano2011,PrassidesNat2010,hirschFullerBook} and carbon nanotubes (nanotubides)\cite{SzirmaiPRB2017}. This route is promising for doping graphene with light alkali atoms and it could address the value of spin lifetime of itinerant electrons and whether Na can differentiate between mono and few-layer graphene, as the latter is known to be inevitably present in chemically exfoliated graphene\cite{SzirmaiSciRep2019}.

Here, we report synthesis of Li and Na doped few-layer graphene (FLG) using liquid ammonia. The FLG material was prepared by chemical exfoliation. Raman spectroscopy, conductivity, and electron spin resonance studies indicate a successful doping. The strongly Fano-like Raman lineshapes evidence a sizable electron-phonon coupling. However, neither SQUID magnetometry nor conductivity measurements gave evidence for a superconducting phase down to $2$ K. The spin-relaxation lifetime of conduction electrons is in excess of $7$ ns which represents an ultralong value especially for chemically exfoliated graphene. We argue that Na dopes selectively monolayer graphene as a few-layer graphene resembles graphite for which Na doping is known to be impossible.

\begin{figure*}[ht!]
	\includegraphics[width=0.7\linewidth]{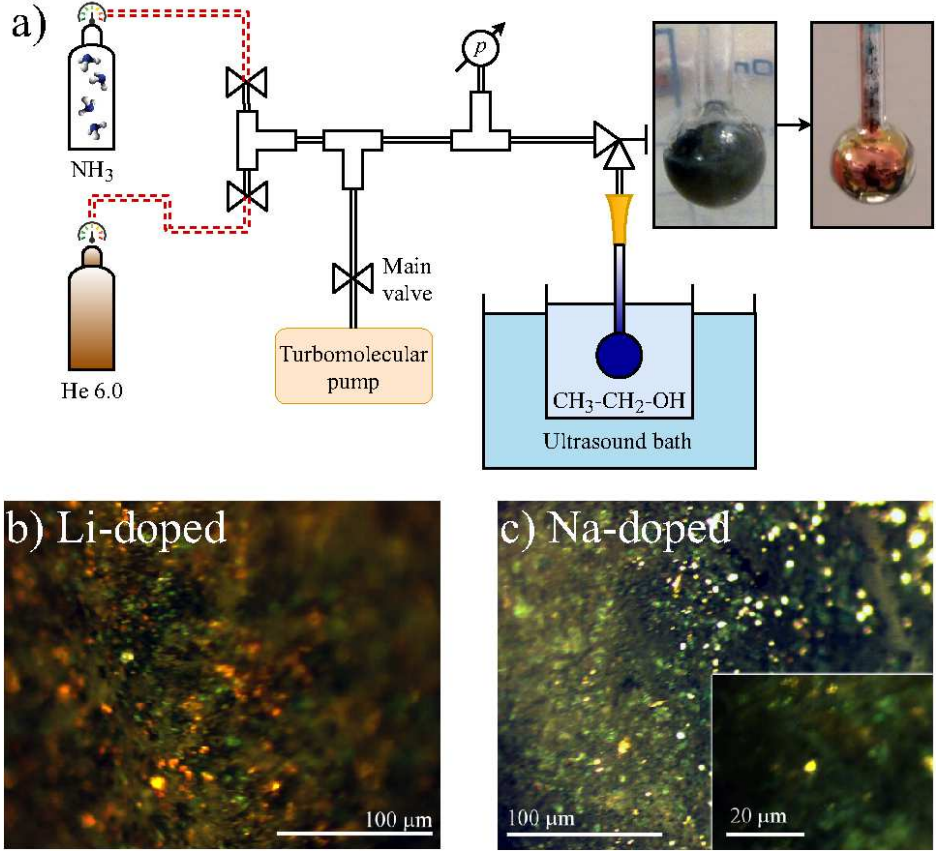}
	\caption{a) Schematics of the doping setup connected to a vacuum line with gas inlets. The alkali-ammonia solution (blue bubble) is kept around $-40~^{\circ}$C using an ethanol bath. A bath sonicator assists the doping and it surrounds the ethanol container. Inset shows the photograph of the solution at the beginning and at the end of the process. b) and c) Microscope images of the final materials. Note that Li doping results in bright, yellow-brownish and bluish flakes, while upon Na doping only bluish areas are present. Inset in c) shows a smaller area of the sodium doped sample shot with a $50\times$ objective to emphasize the presence of bluish areas.}
	\label{fig:Fig1_preparation}
\end{figure*}

\section{RESULTS AND DISCUSSION}

Few-layer graphene (FLG) samples were doped with lithium and sodium in liquid ammonia with the setup shown in Fig. \ref{fig:Fig1_preparation}a. We previously demonstrated that the starting material mainly consists of graphene flakes with $5$ or less layers with a log-normal distribution centered around $3$ layers \cite{SzirmaiSciRep2019}. Photographs of the samples are presented in Fig. \ref{fig:Fig1_preparation}b. and c. A characteristic color change was observed to the black FLG material upon doping, which is a fingerprint of charge transfer from the alkali atoms toward the carbon, as seen before for graphite intercalation compounds\cite{DresselhausAP1981,ChaconPRB2012} and graphene \cite{JungACSNano2011,HowardPRB2011,SzirmaiSciRep2019}. In case of lithium, both yellow-brownish and bluish spots are present, which attests a successful doping of the material. For sodium, only bluish areas are observable, which hints at lighter, yet successful doping. For intercalated graphite, yellow color is the signature of the highest level of doping (also referred to as stage-I compounds), whereas blue color is a signature of a lower level of doping \cite{DresselhausAP1981}. 

The most striking observation of this synthesis process is the presence of \emph{any} color change to the Na doped FLG sample. Na is known to intercalate (or dope) graphite  to a very low level such as NaC$_{64}$ \cite{AsherNat1958,AsherJINC1959,MetrotSM1980,AkuzavaMCLC2002} which has the same color as graphite. In fact, higher level of Na doping can only be achieved with an ultra high pressure synthesis method \cite{RussianHighP_NaDoping} but the resulting material is unstable at ambient pressure. It has long been intriguing why Na does not effectively dope graphite. Graphite intercalation compounds possess a long-range ordered structure for the alkali atoms and also for the graphene layers which are adjacent to layers of alkali atoms. Recent \emph{ab initio} calculations \cite{Adelhelm_why_NaGIC_unstable} hint at a delicate interplay between the alkali ion induced structural deformation (and the corresponding weakening in the van der Waals interaction between graphene layers), the covalent (for Li) \textit{versus} ionic (for Na) interaction which makes the formation of Na intercalated graphite thermodinamically unfavored. In contrast, sodium can effectively dope fullerides up to Na$_4$C$_{60}$ \cite{OszlanyiPRL1997,BrouetPRL2001,BrouetPRB2002}, where the voids between the fullerene balls can be conveniently filled thus the deformation energy of the van der Waals molecular crystal is smaller and is compensated by the energy gain during the electron charge transfer. Therefore it is expected that Na could dope monolayer graphene where the energy loss due to deformation is also absent.

\begin{figure}[h!]
\includegraphics[width=1.0\linewidth]{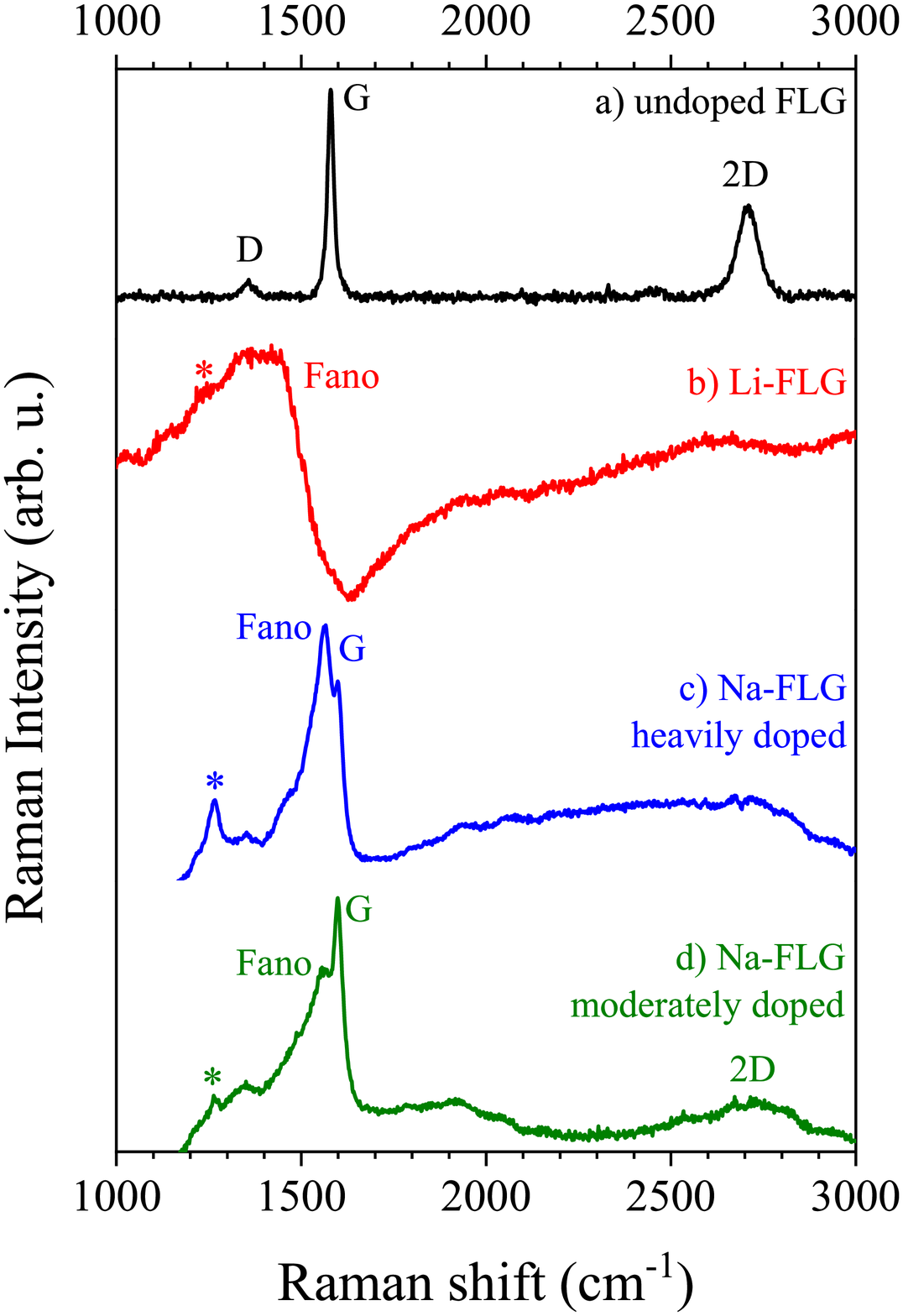}
\caption{Raman spectra taken at $514.5$ nm wavelength. a) Undoped FLG shows the usual D, G and 2D modes. b) A broad and intensive Fano line at $1510$ cm$^{-1}$ dominates the spectrum in the lithium doped FLG sample; the D and 2D modes are barely visible and the latter is extremely broadened. c)-d) Sodium doped samples also display a Fano line at $1566$ cm$^{-1}$ and a symmetric peak at $1600$ cm$^{-1}$. The latter is denoted by G and is associated with weakly charged graphene flakes. Asterisk denotes a peak from the quartz sample holder.}
\label{fig:Fig2_raman514}
\end{figure}

Given the surprising result of a successful and relatively high level of Na doping of the few layer graphene sample, we performed Raman spectroscopy on the synthesized materials as it is a sensitive probe of the charge transfer toward the carbonaceous material. Raman spectra recorded at $514.5$ nm wavelength of the materials are presented in Fig. \ref{fig:Fig2_raman514}. Raman spectrum of the pristine material displays the usual D ($1358$ cm$^{-1}$), G ($1580$ cm$^{-1}$), and 2D ($2710$ cm$^{-1}$) bands and it reproduces the earlier reports on similar samples \cite{EnglertNatChem2011}. 

The lithium doped FLG sample shows a broad and intensive Fano-shaped line (also referred as Breit-Wigner-Fano or BWF lineshape) \cite{FanoPR1961,KuzmanyBook} at $1507$ cm$^{-1}$, which dominates the spectrum. The D and 2D modes are barely visible and the latter is extremely broadened. The Fano line is downshifted and flattened compared to the original G mode, as expected upon heavy doping. Compared to the Raman spectrum of LiC$_6$ GIC, the observed peak is relatively close to its E$_{2g_{1}}$ mode at $1546$ cm$^{-1}$. The doped samples are in a quartz tube which explains the origin of the Raman line denoted by an asterisk.

We also present Raman data for two Na doped samples, one denoted as heavily, the other as moderately doped. The heavy doping notation refers to the highest achievable doping level whereas the moderate doping is representative for a sample prepared with a smaller amount of alkali dopants. Both types of sodium doped samples present a Fano line at around $1572$ cm$^{-1}$ and a symmetric Lorentzian peak at $1600-1602$ cm$^{-1}$. The Fano peak at the lower Raman shift confirms the presence of highly intercalated flakes in agreement with the above mentioned surprising microscopic observation, \textit{i.e.} that significant charge transfer is observed in the Na doped samples. {\color{black} The Fano lineshape in alkali-metal intercalation compounds with graphite only occurs when a stage-I compound is reached, that means when each graphene layer is surrounded from both sides by an alkali atom layer \cite{ChaconACSNano2013}. The observation of this asymmetric mode in the Na doped sample provides a direct proof for the existence of Na doped monolayer graphene.} The symmetric peak at higher Raman shifts (denoted by G) is associated with weakly charged or incompletely intercalated flakes \cite{AkuzavaMCLC2002}. The 2D mode is very broad in all samples as expected for a high level of doping \cite{ChaconPRB2012,ChaconACSNano2013}. 

In general, the Fano lineshape is the result of quantum interference between the zone-center phonons and the electronic transitions \cite{FanoPR1961,KuzmanyBook} thus it is an important benchmark of a significant charge transfer in carbonaceous materials including fullerides \cite{KuzmanyAdvMater1994}, carbon nanotubes (nanotubides) \cite{RaoNat1997}, graphite \cite{EklundPRB1977}, and graphene (graphenides) \cite{JungACSNano2011,HowardPRB2011,ParretACSNano2013}. 

Additional information about the electronic structure, the level of charge transfer, and the magnitude and sign of the electron-phonon coupling can be obtained from the details of the Fano lineshape. According to Refs. \cite{FanoPR1961,KuzmanyBook}, the lineshape reads as a function of the Raman shift energy, $\hbar \omega$:

\begin{equation}
I(\omega)= I_0 \frac{\left(q+\overline{\epsilon}\right)^2}{1+\overline{\epsilon}^2}+A,
\end{equation}
where $q$ is the Fano asymmetry lineshape parameter, $\epsilon=\frac{\omega-\omega_0-\Delta}{\Gamma/2}$ and $A$ is an offset parameter. Herein $\hbar \omega_0$ is the vibrational energy of the unperturbed phonon, $\hbar\Delta$ is the interaction induced shift of the phonon energy and $\hbar \Gamma$ is the broadening parameter due to the interaction. In general $\Gamma$ and $\Delta$ increases with the strength of the electron-phonon coupling however $q$ is inversely proportional to it, $q=\pm \infty$ represents the non-interacting limit and $q \rightarrow 0$ is the limit of strong electron-phonon interaction. 

We found $q=-1.03$ for the Li doped FLG and $q=-2.64$ and $q=-3.17$ for the heavily and moderately Na doped FLG samples, respectively (details of the lineshape analysis are provided in the Supplementary Information). These figures have to be compared with $q=-1.09$ found for the graphite intercalation compounds \cite{ChaconPRB2012} LiC$_6$, KC$_8$, and CaC$_6$, of which the latter is a superconductor with $T_{\text{c}}=11.5\,\text{K}$ (Ref. \cite{CaC6NatPhys}). 



The significant charge transfer and the sizable electron-phonon coupling in the Li doped FLG motivated us to search for traces of superconductivity for both the Li and Na doped materials. However, SQUID magnetometry in a zero field cooled condition and using a 10 Oe of sensing field, while warming up the samples from 2 K, did not reveal any traces of superconductivity.

\begin{figure}[h!]
\includegraphics[width=1.0\linewidth]{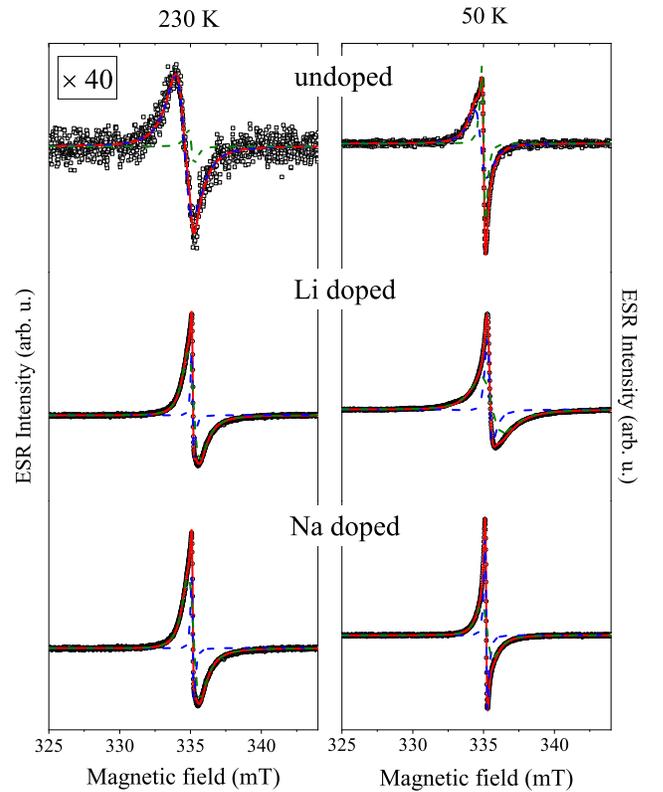}
\caption{ESR spectra of the undoped, Li and Na doped FLG samples at $230$ K and $50$ K. Open symbols are the measured data, the fitted curve is represented with a continuous red line. Blue and green curves are the decomposition to the two components present for each sample. The two lines are fully symmetric in the undoped material, whereas both components are asymmetric in the doped ones, indicating a metallic sample. Note the zoomed vertical scale for the undoped material.}
\label{fig:esr_spectra}
\end{figure}

Electron spin resonance spectroscopy \cite{Slichter} can provide a great deal of information about the electrons in a material, including localized electron spins, which are related to defects or dangling bonds, and also about delocalized, \textit{i.e.} conduction electron species.
Fig. \ref{fig:esr_spectra}. presents the ESR spectra of the undoped and Li, Na doped FLG samples at $230$ K and $50$ K. In each cases, the observed lineshape can be decomposed into two separate lines. In the undoped material, the lines are completely symmetric and thus can be fitted with derivative Lorentzians. The origin of these lines is most probably dangling bonds or lattice defects, which give rise to a paramagnetic signal. The presence of such lines is very common in carbon materials including fullerenes \cite{NemesPRB2000}, carbon nanotubes and graphite \cite{GalambosPSSB2009}, graphene \cite{MarkusPSSb2015} or in boron doped diamond \cite{SzirmaiPRB2013}.

Upon doping the FLG sample with Li and Na, the ESR spectra change significantly: strong signals with about 30-50 times larger intensity appear with an asymmetric lineshape. An asymmetric ESR lineshape is identified as a so-called Dysonian line \cite{DysonPR1955}, which is the usual case in metallic materials. In fact, the Dysonian lineshape in ESR can be considered as analogue to the Fano lineshape in Raman spectroscopy, however its physical origin is different. The Dysonian lineshape is due to the varying microwave phase along the sample volume due to the limited penetration of microwaves. The asymmetry is stronger for both lines in the lithium doped material than in the sodium doped one. This indicates a smaller microwave penetration depth thus a larger conductivity. This observation is in agreement with the visual observations and the Raman spectroscopy results above, \textit{i.e.} that charge transfer is stronger in the lithium doped sample. 

Fig. \ref{fig:esr_spectra}. shows that the ESR signal in the alkali doped samples can be decomposed into two distinct ESR lines with different linewidths. The presence of multiple ESR lines is often encountered in alkali atom doped carbon materials. The examples include alkali doped fullerides  \cite{JanossyPRL1993,NemesPRB2000} and carbon nanotubes (nanotubides)\cite{ClayePRB2000,SzirmaiPRB2017}. The origin of multiple ESR lines in doped carbon could be an inhomogeneous doping, or the presence of localized paramagnetic spins.

To gain a deeper insight to the electronic properties of the doped materials, the temperature dependence of the ESR signal was studied in the $5$ to $250$ K temperature range. The ESR intensity is directly proportional to the spin susceptibility, which allows to identify the nature of spins (localized or delocalized) which give rise to the ESR signal. In addition, the ESR linewidth is related to $T_2$, which is called the spin-spin relaxation time (due to historical reasons) \cite{DysonPR1955,FeherPR1955,Slichter}. In the following, we refer to this as spin-relaxation time or $\tau_{\text{s}}$ to conform with the spintronics literature. Determining $\tau_{\text{s}}$ is directly relevant for the spintronics applications of graphene.

The spectroscopic properties, including line intensity, position, and linewidth were determined by fitting the Dysonian lines with a mixture of absorption and dispersion Lorentzian lines \cite{WalmsleyJMR1996,DjokicJMMM2019}, which is a valid approach when the conduction electrons are diffusing through the microwave penetration depth slowly (this is the so-called NMR limit) compared to their lifetime \cite{DysonPR1955}. 


\begin{figure}[h!]%
	\includegraphics[width=1.0\linewidth]{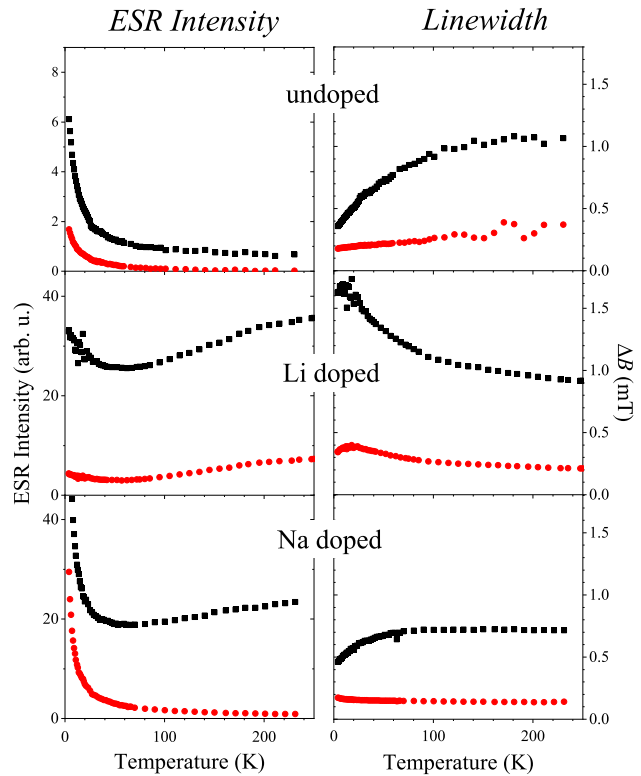}
	\caption{Temperature dependent ESR intensity (normalized by the sample mass) and linewidth. Note the significantly smaller scale for the undoped material. The lithium doped sample shows two Pauli-like signal intensity, which is a clear indication of a metallic behavior. On the other hand, the sodium doped material shows a mixed behavior with one Curie and one Pauli-like signals.}
	\label{fig:esr_int_lw}
\end{figure}

The temperature dependent ESR intensity normalized by the sample mass and the ESR line-width is shown in Fig. \ref{fig:esr_int_lw}. It reinforces the earlier observation that the observed spin susceptibility increases significantly upon doping. In addition, the temperature dependent character of the spin susceptibility also characteristically changes: in the undoped sample, the spin susceptibility shows a significant increase upon lowering temperature, which is characteristic for localized, \textit{i.e.} paramagnetic spin species. We thus term it a "Curie-like" behavior, where $\chi_{\mathrm{s}} \sim 1/T$. In fact, the intensity of the broader peak in the undoped material remains finite at higher temperature. This can be related to an unintentional doping due to some residual solvent or contamination in the starting material.

In contrast, a Pauli-like spin susceptibility, \textit{i.e.} with little temperature dependence, dominates the signal for both the lithium and sodium doped samples above 20 K, although the sodium doped material also contains a Curie-like contribution with a much smaller intensity. An ESR signal with a spin-susceptibility with little or no temperature dependence is characteristic for conduction electrons. The fact, that the Li doped material contains two such signals, hints that the doping is inhomogeneous. We note that the presence of unreacted metallic particles can be excluded as the origin of the observed metallic signals, since metallic Li and Na have a characteristic lineshapes (typical for strongly diffusing electrons) which is not observed herein \cite{FeherPR1955,DysonPR1955}. Even the Pauli-like signal shows an upturn in intensity in the Na doped sample below 20 K; similar effects are often encountered when a minute amount (of the 10 ppm level/lattice sites) of paramagnetic spins give a common resonance with the itinerant electrons \cite{BarnesAdvPhys1981}. This also means that the linewidth data in the Na doped is only reliable above 20 K.

We identify the Pauli-like signal in the Na doped FLG sample as coming from Na doped graphene monolayers. It is based on the observation that Na cannot dope graphite, thus it is very probable that multilayer graphene is also inaccessible for it. This observation is in a full agreement with the Raman result, as it evidenced that Na doping of an FLG sample leaves a part of the sample undoped. 


\begin{figure}[h!]%
	\includegraphics*[width=1.0\linewidth]{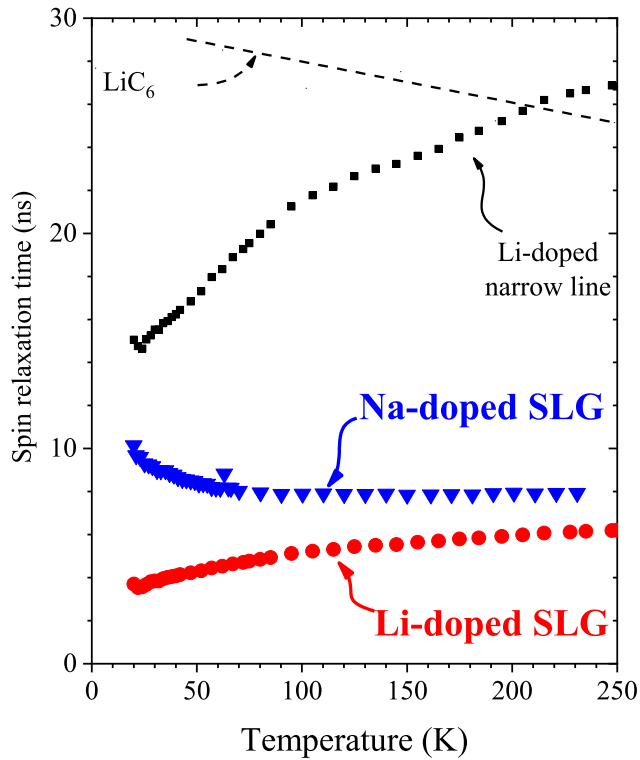}
	\caption{Spin relaxation times of the Li and Na doped few layer graphene above 20 K. Dashed line shows data on the LiC$_6$ intercalated graphite powder from Ref. \cite{LauginiePhys1980}.}
	\label{fig:esr_relax}
\end{figure}

The ESR linewidth, $\Delta B$, data in Fig. \ref{fig:esr_int_lw} can be used to directly obtain the spin relaxation time through $1/\tau_{\text{s}}=1/\gamma \Delta B$, where $\gamma=2\pi\cdot 28.0\,\text{GHz/T}$ is the electron gyromagnetic factor. The result is shown in Fig. \ref{fig:esr_relax}. along with a reference data from LiC$_6$ intercalated graphite powder from Ref. \cite{LauginiePhys1980}. The narrow component of the Li doped FLG sample has a relaxation time that is similar to that in LiC$_6$. It is therefore tempting to identify this component as coming from Li intercalated graphene multilayers and the broader component to Li doped monolayers. The spin relaxation time of this component is similar to that found for the Na doped graphene monolayers. 

In principle, the temperature dependent character of the spin relaxation time also contains information about the relaxation mechanism\cite{ZuticRMP2004,ZuticBook2012}, whether it can be described by the Elliott-Yafet \cite{ElliottPR1954,YafetSSP1963} or the D'yakonov-Perel' theories \cite{DyakonovPerel}. However, a proper theoretical description requires the knowledge of the temperature dependent mobility and the local crystalline structure in these samples. Neither of these are available at present, therefore we cannot speculate on the underlying spin-relaxation mechanism.

Nevertheless, the striking observation is that the ESR linewidth at 230 K, when translated to spin-relaxation times, gives $\sim 6.1$ ns and $\sim 7.9$ ns for lithium and sodium doped monolayer graphene, respectively. 
These values are even more surprising in view of the well-known heterogeneity of chemically exfoliated graphene and the synthesis method we employed. In fact, our figures are only surpassed by recent spin-transport experiments on carefully manufactured microscopic graphene samples, with ultrahigh mobility.

Early spin-transport experiments reported sub-nanosecond spin-relaxation times in graphene \cite{TombrosNAT2007,HanPRL2010,YangPRL2011,HanPRL2011,OzyilmazNL2011,HanNL2012,VolmerPRB2013}, which most probably originates from extrinsic effects \cite{KochanPRL2014}. Advanced sample preparation and coating with hexagonal-BN resulted in improved values \cite{DrogelerNL2014,GuimaraesPRL2014} with the longest spin-relaxation time values being 12.6 ns \cite{DrogelerNL2016} in single-layer graphene and 9.4 ns in bilayer-graphene \cite{LeutenantsmeyerPRL2018}. 

It is in fact astonishing that a chemically prepared sample, which is readily available in mg quantities, displays the same magnitude for the spin-relaxation time as the above values which are the result of advanced micromanipulation techniques, combined with van der Waals heterostructure engineering.
The advanced techniques aim to decrease the defect concentration of individual graphene microstructures. However, bulk, top-down chemical methods may lead to higher quality flakes due to a finite probability that the flake defect concentration may even be smaller than those achieved by local techniques. As the sodium doping can only occur in high-quality single flakes, this may provoke a selectivity that further increases the measured spin-relaxation time{\color{black}, thus leading to the efficient manufacturing of future spintronics materials based on graphene}.

\section{CONCLUSIONS}

In conclusion, we successfully synthesized lithium and sodium doped few-layer graphene in liquid ammonia solution. We found that Na dopes exclusively monolayer graphene which is present in the FLG sample. 
 This is deduced from an overally weaker Na doping of the FLG samples, which is argued to be associated with the well-known inability of Na to intercalate graphite, thus multilayer graphene is also inaccessible for it. The obtained materials exhibit a clear change of color, a dominant Fano mode in the Raman spectrum, and ESR signals due to conduction electrons. These observations prove a successful intercalation and charge transfer from the alkali atoms toward graphene. Electron spin resonance spectroscopy indicates a spin-relaxation time of 6-8 ns, which is comparable to the longest values found on ultrahigh mobility graphene flakes using spin-transport methods. Given that we studied a relatively impure and bulk graphene material, these figures {\color{black}are encouraging} for the spintronics applications of alkali atom doped graphene.

\section{METHODS}
\textbf{Preparation of the few layer graphene starting material.}
Few-layer graphene was prepared from saturation potassium doped spherical graphite powder (SGN18, Future Carbon) using DMSO solvent for the wet chemical exfoliation as described previously \cite{EnglertNatChem2011,VeceraPSSb2014,VeceraNatCom2017}. Chemical exfoliation was finalized using ultrasound tip-sonication, which yields the best quality, as shown in a previous study \cite{MarkusPSSb2015}. The properties of the starting material are well characterized by atomic force microscopy and Raman spectroscopy, which revealed that restacked few-layer graphene is also present in the sample\cite{SzirmaiSciRep2019}.
Prior to intercalation, the undoped FLG was heated up to $400~^{\circ}$C for $30$ minutes in high vacuum ($2\times10^{-6}$ mbar) to remove any residual solvents. It was shown previously Refs. \cite{MarkusPSSb2015,SzirmaiSciRep2019} that this does not affect the morphology of the starting FLG.

\textbf{Doping of graphene with lithium and sodium}
{\color{black} Lithium and sodium with a purity of $99.9+\%$ and $99.8\%$ (from Sigma-Aldrich), respectively, were handled in an Ar glove box (O$_2$, H$_2$O~\textless~0.1 ppm). The lithium granules had metallic color and were cut into smaller pieces to increase the surface. The sodium chunks were first thoroughly cleaned and then cut up. About $1$ milligram of FLG and excess alkali metal of $1.1$ or $2.2$ milligrams for lithium and sodium, respectively, were placed in a bulb shaped quartz container (reaction chamber in the following) inside the glove box. The bulb of the reaction chamber (hand-blown by a technician) had a diameter of 10 mm and it was connected to the vacuum line with a 5 mm diameter quartz tube. A (Li,Na)C$_6$ stoichiometry (which is probably the upper limit of doping) would require 0.09 mg and 0.24 mg for the Li and Na, respectively, for the 1 mg FLG. The unreacted mixture was then connected to a vacuum line with the help of a well-sealing valve and pumped to a vacuum better than $10^{-6}$ mbar.

The reaction chamber was placed in liquid nitrogen and gaseous ammonia, that was connected to the vacuum line, was rapidly (within a few seconds) condensed to the starting mixture of FLG and alkali metal. The rapid condensing is required to prevent the formation of side products. Following this, the reaction chamber is surrounded by an ethanol bath which is kept at $-40~^{\circ}$C (below the $-33~^{\circ}$C boiling point of ammonia at ambient pressure). Ethanol has a freezing point of $-114~^{\circ}$C thus it is well suited for this purpose. The reaction chamber, surrounded by the ethanol bath was immersed in a bath sonicator where it was intensively sonicated for $30$ minutes to obtain a homogeneous doping of the few-layer graphene. The pressure of the chamber is monitored to avoid evaporation of the ammonia during the process. The dissolution of alkali metals in liquid ammonia can be followed by characteristic color changes: the solution changes its color to dark blue initially, immediately followed by a change to yellow-brownish. This marks the dissolution of the alkali metals: dark blue color indicates a low concentration of the dissolved alkali and yellow-brownish color is characteristic for a high concentration of dissolved alkali metals \cite{greenwood1998}. Finally, the reaction chamber changes its color to a homogeneous brown which indicates the doping of FLG with the alkali metals.

At the end of the intercalation, the solution is slowly heated to room temperature, when ammonia is evacuated. Subsequently, a heating of $200~^{\circ}$C is applied for $30$ minutes to remove any absorbed ammonia and the unreacted alkali metal. The choice of  $200~^{\circ}$C is delicate: on one hand it is above the melting temperature of the alkali metals ($181~^{\circ}$C for Li and $98~^{\circ}$C for Na) which allows for an efficient removal of the unreacted, excess alkali atoms but on the other hand, it is below the onset of the formation temperature of alkali carbides (this occurs above $\sim 400 ~^{\circ}$C, Ref. \cite{BasuMSE1979}). The reaction chambers were sealed off with a torch on the 5 mm quartz tube part.} 


Ammonia is known to react with alkali metal elements and forms amides \cite{BergstromCR1933,JuzaAngew1964}. This side reaction is known to proceed fast when the ammonia is in the gas phase. However, this reaction channel is prohibited (or slowed down to the week timescale), when the ammonia is liquefied. Besides, the amides have well-known signatures in Raman spectroscopy \cite{CunninghamJCP1972,LiuJPCB2011} in the form of a series of lines around $400-550$ cm$^{-1}$. These signals were used to to monitor the content of this unwanted side product. We optimized the procedure to yield a material with no traces of amides \cite{MarkusPSSb2019}.
We emphasize at this point that while we refer to the material as "Na doped FLG", Na dopes exclusively the monolayer graphene content inside the sample.

\textbf{Raman spectroscopy.}
{\color{black}Raman spectroscopy was performed on the as prepared samples inside the same reaction chambers where the doping proceeds.} {\color{black}A $514.5$ nm wavelength laser excitation with $0.5$~mW power was used to avoid laser-induced deintercalation~\cite{NemanichPRB1977,ChaconPRB2012} or sample heating.} Raman spectra were recorded on a modified broadband LabRAM spectrometer (Horiba Jobin-Yvon Inc.). The built-in interference filter was replaced by a beam splitter plate with $30\%$ reflection and $70\%$ transmission to allow for a broadband operation \cite{FabianRSI2011,FabianPSSB2011}. Typically $0.5$~mW laser powers were used with a built-in microscope (Olympus LMPlan $50\times/0.50$ inf$./0/$NN$26.5$), which yields about $1\mu\mathrm{m} ~\times 1\mu\mathrm{m}$ spot size. Photographs were taken with an Olympus $10\times~/~50\times$ objective, which is present on the LabRAM equipment.

\textbf{Electron spin resonance}
{\color{black}The reaction chambers were opened up inside an Ar glove box and placed into quartz tubes with a 4 mm outer diameter. These tubes were then evacuated on the vacuum line to better than $2\times10^{-6}$ mbar, then filled with 20 mbar He exchange gas for the cryogenic measurements and sealed permanently with a torch. Raman spectrum of the materials were checked before and after this procedure to ensure that no changes occur during this operation.} ESR measurements were performed on a Bruker Elexsys E500 X-band spectrometer equipped with an Oxford He flow cryostat. The temperature could be varied between $4$ and $300$ K. Care was taken to avoid saturation and overmodulation of the observed signals, similarly to our previous works \cite{SzirmaiPRB2017,SzirmaiPSSB2011}.  The spectral parameters of each signal component is determined by fitting (derivative) Lorentzian and Dysonian curves, as is customary in the ESR literature. Due to the magnetic field modulation technique employed in ESR, the observed lineshapes are the derivative of the following (with respect to the magnetic field, $B$):

	\begin{align}
	I(B)=A \cdot \frac{\Delta B \cos\left(\phi \right)+\left(B-B_0\right)\sin\left(\phi \right)}{\left(B-B_0 \right)^2+\Delta B^2},
	\end{align}

\noindent where the ESR lineshape, $I(B)$ is given as a function of the $B$ magnetic field, $A$ is proportional to the static spin susceptibility, $B_0$ is the line position, $\Delta B$ is the ESR linewidth, $\phi$ is the mixing angle between the two types of curves (absorption and dispersion). The ESR linewidth is related to the $T_2$ spin-spin relaxation time as $\Delta B=1/\gamma T_2$ where $\gamma=2\pi\cdot 28.0\,\text{GHz/T}$ is the so-called electron gyromagnetic ratio. 

\textbf{Magnetometry with a superconducting quantum interference device}
{\color{black}Samples inside the same, He exchange gas filled quartz tubes were used for the SQUID measurement that were used in the ESR studies.} We searched for traces of superconductivity in a standard MPMS SQUID magnetometer down to $2$ K with a zero magnetic field cooling protocol, in small applied fields of $5-50$ Oe, and also at $2$ K in magnetic hysteresis loops.

\section{ASSOCIATED CONTENT}

\textbf{Supporting Information}\\
The Supporting Information is available free of charge at https://pubs.acs.org/doi/..............\\
\smallskip

Details of the Raman spectra deconvolution, spectroscopic parameters of the Breit-Wigner-Fano lineshapes, a discussion of the electron-phonon coupling and the temperature dependent $g$-factor as detected by electron spin resonance (PDF).

\section{AUTHOR INFORMATION}
\textbf{Author Contributions}\\
B.G.M. performed the doping, Raman and ESR studies under the supervision of F.S., P.Sz., B.N. and L.F. contributed to the Raman and ESR investigations. K. E., P. E. prepared the starting FLG samples under the supervision of A. H and F. H. SQUID measurement were performed by N. M. N. Raman studies and the analysis was performed by J.C.C-T. and T. P. All authors contributed to the writing of the manuscript.

\smallskip

\noindent\textbf{Notes}\\
The authors declare no competing financial interest.

\section{ACKNOWLEDGEMENTS}
{\color{black}The Authors are indebted to Andr\'as Magyar for the preparation of the reaction chambers.} We wish to kindly thank Prof. J.~L.~Martinez for performing the SQUID measurements. Support by the National Research, Development and Innovation Office of Hungary (NKFIH) Grant Nrs. K119442 and 2017-1.2.1-NKP-2017-00001 are acknowledged. T. P. thanks the FWF (P27769-N20) for funding.



\clearpage
\newpage

\section*{SUPPLEMENTARY MATERIAL}
\setcounter{section}{1}
\setcounter{figure}{0}
\makeatletter 
\renewcommand{\thefigure}{S\@arabic\c@figure} 

\section{Deconvolution of the Raman spectra}

\begin{figure}[h!]%
	\includegraphics*[width=0.7\linewidth]{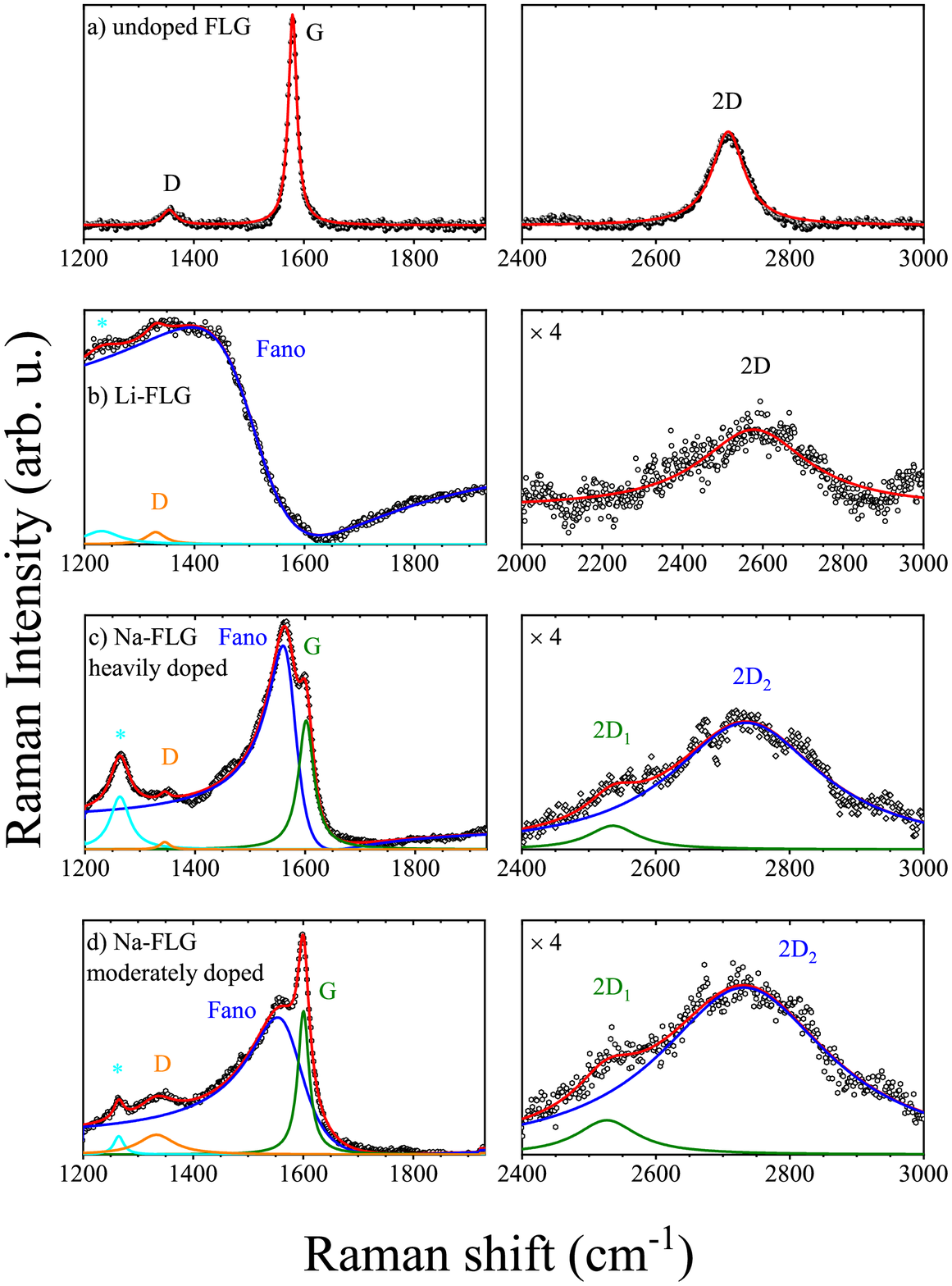}
	\caption{Deconvolution of the observed Raman spectra for the various samples.}
	\label{fig:raman_deconv}
\end{figure}

Deconvolution of the observed Raman spectra is depicted in Fig. \ref{fig:raman_deconv}. The sum of the lines are shown in red. On the left side, the asymmetric Fano lineshape is denoted with blue color. The peak only present around $1600$ cm$^{-1}$ in the sodium species are green. The D mode has an orange color in the figure. The parasitic signal arising from the quartz tube is marked with $\ast$ and shown in cyan color. On the right side, the extremely broadened 2D mode is shown. For the case of lithium it composes only a single Lorentzian, while for sodium it splits into two. This is probably related to the argument that only the monolayer flakes get truly charged.

We note that in LiC$_6$ (Ref. \cite{ChaconPRB2012}), the transformation of the G mode to a dominant Fano lineshape is accompanied by the complete disappearance of the 2D mode, which is not the case herein. This is probably related to the sample heterogeneity and the resulting inhomogeneous doping in agreement with Fig. 1b. of the main text, where the characteristic red/golden spots of a highly intercalated Li compound are not homogeneously dispersed along the sample. Also, we cannot exclude that besides the dominant Fano line, a small G mode (corresponding to un- or weakly doped graphene layers) exists, whose presence is suggested from the strongly broadened 2D-mode observed in Fig. \ref{fig:raman_deconv}b., which in single LiC$_6$ crystals is not present \cite{ChaconPRB2012}.

The obtained parameters from the fit of the G mode related peaks are summarized in Table \ref{tab:raman_params}. The complete deconvolution of the observed peaks are visualized in Fig. \ref{fig:raman_deconv}. The electron-phonon coupling parameter is calculated from the fitted values, as described later.

\begin{table*}[h!] 
	\begin{center}
		\begin{tabular*}{\linewidth}{c @{\extracolsep{\fill}} cccccc} \hline \\
			Sample       & $\omega_{\mathrm{G}}$ & $\Gamma_{\mathrm{G}}$ & $\omega_{\mathrm{Fano}}$ & $\Gamma_{\mathrm{Fano}}$ & $q$     & $\gamma^{\mathrm{EPC}}$ \\ \\ \hline
			Undoped      & $1580$                & $10$                  &                          &                          & \\
			Li-doped     &                       &                       & $1507$                   & $116$                    & $-1.03$ & $206$ \\
			LiC$_6$ [Ref. \cite{ChaconPRB2012}]& $1585$ & $71$                  & $1546$                   & $71$                     & $-1.09$ & $157$ \\
			Na-doped (heavily)& $1602$           & $16$                  & $1572$                   & $31$                     & $-2.64$ & \\
			Na-doped (moderately)& $1600$        & $16$                  & $1572$                   & $61$                     & $-3.17$ & \\ \hline
		\end{tabular*}
		\caption{Obtained parameters of the fitted Raman modes. The position and width values are in cm$^{-1}$.}
		\label{tab:raman_params}
	\end{center}
\end{table*}

To obtain further valuable information of the nature lithium intercalation, the electron-phonon coupling parameter (EPC), which is directly related to the charge transfer can be calculated from the the fitted parameters, whose are summarized in Table \ref{tab:raman_params} and Figure \ref{fig:raman_deconv} of Supplementary Materials, using:

\begin{equation}
\gamma^{\mathrm{EPC}}=2\sqrt{(\omega_{\mathrm{Fano}} - \omega_{\mathrm{A}})(\omega_{\mathrm{NA}}-\omega_{\mathrm{Fano}})}.
\end{equation}
Here, $\omega_{\mathrm{Fano}}$ is the measured position of the G-line peak, $\omega_{\mathrm{A}}$ and $\omega_{\mathrm{NA}}$ are the calculated adiabatic and non-adiabatic phonon frequencies \cite{SaittaPRL2008}. We approximate the latter two quantities with the ones calculated for LiC$_6$: $\omega_{\mathrm{A}}=1362$ cm$^{-1}$ and $\omega_{\mathrm{NA}}=1580$ cm$^{-1}$, as no exact calculation exists for FLG. This approximation was found to be valid in similar hexagonal carbon systems such as potassium doped multiwalled carbon nanotubes \cite{ChaconTorresCarb2016}. The calculated EPC value for the lithium doped sample is $\gamma^{\mathrm{EPC}} = 206$ cm$^{-1}$, which is similar to that of LiC$_6$, where it is $157$ cm$^{-1}$ \cite{ChaconPRB2012}. Unfortunately, no adequate calculation exists for the case of sodium.

\section{Change of $g$-factor upon alkali intercalation}

\begin{figure}[h!]
	\includegraphics*[width=0.7\linewidth]{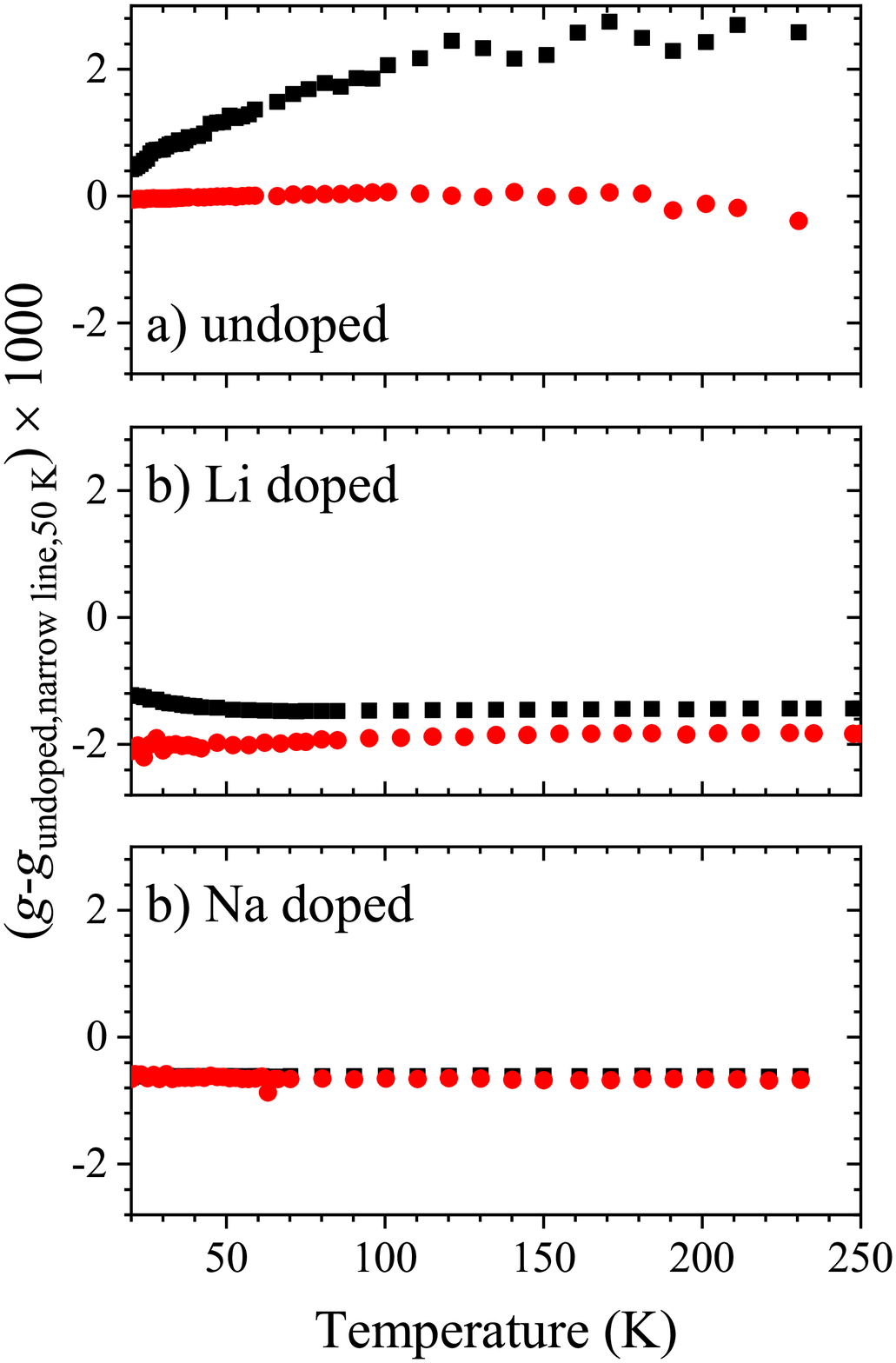}
	\caption{Relative $g$-factor values compared to $g_{\mathrm{undoped},\mathrm{narrow line},50~\mathrm{K}}=2.0069$, which corresponds the narrow line of the undoped material at $50$ K. Please note, that both lines in the intercalated materials are downshifted, which is expected according to the Elliott-Yafet theory \cite{ElliottPR1954,YafetSSP1963}.}
	\label{fig:esr_gfactor}
\end{figure}

ESR $g$-factors of the investigated materials are shown in Fig. \ref{fig:esr_gfactor}. with respect to that of $g_{\mathrm{undoped},\mathrm{narrow line},50~\mathrm{K}}=2.0069$ of the narrow line of the undoped material at $50$ K. Please note that each line present in the intercalated materials are downshifted relative to this, which is an expected behavior upon $n$ doping, according to the Elliott--Yafet theory \cite{ElliottPR1954,YafetSSP1963}. This is also a clear indication of charge transfer and successful intercalation.


\providecommand{\latin}[1]{#1}
\makeatletter
\providecommand{\doi}
{\begingroup\let\do\@makeother\dospecials
	\catcode`\{=1 \catcode`\}=2 \doi@aux}
\providecommand{\doi@aux}[1]{\endgroup\texttt{#1}}
\makeatother
\providecommand*\mcitethebibliography{\thebibliography}
\csname @ifundefined\endcsname{endmcitethebibliography}
{\let\endmcitethebibliography\endthebibliography}{}

\end{document}